# An IoT Platform-as-a-Service for NFV Based – Hybrid Cloud / Fog Systems

Carla Mouradian[¥], Fereshteh Ebrahimnezhad[¥], Yassine Jebbar[¥], Jasmeen Kaur Ahluwalia[¥], Seyedeh Negar Afrasiabi[¥], Roch H. Glitho[¥], Ashok Moghe[€]

[¥]CIISE, Concordia University, Montreal, QC, Canada,
[€]CISCO Systems, San Francisco Bay Area, USA

*Abstract* — **Cloud computing, despite its inherent advantages (e.g., resource efficiency) still faces several challenges. The wide area network used to connect the cloud to end-users could cause high latency, which may not be tolerable for some applications, especially Internet of Things (IoT) applications. Fog computing can reduce this latency by extending the traditional cloud architecture to the edge of the network and by enabling the deployment of some application components on fog nodes. Application providers use Platform-as-a-Service (PaaS) to provision (i.e., develop, deploy, manage, and orchestrate) applications in cloud. However, existing PaaS solutions (including IoT PaaS) usually focus on cloud and do not enable provisioning of applications with components spanning cloud and fog. Provisioning such applications requires novel functions, such as application graph generation, that are absent from existing PaaS. Furthermore, several functions offered by existing PaaS (e.g., publication/discovery) need to be significantly extended in order to fit in a hybrid cloud/fog environment. In this paper, we propose a novel architecture for PaaS for hybrid cloud/fog system. It is IoT use case-driven, and its applications' components are implemented as Virtual Network Functions (VNFs) with execution sequences modeled as graphs with sub-structures such as selection and loops. It automates the provisioning of applications with components spanning cloud and fog. In addition, it enables the discovery of existing cloud and fog nodes and generates application graphs. A proof of concept is built based on Cloudify open source. Feasibility is demonstrated by evaluating its performance when PaaS modules and application components are placed in clouds and fogs in different geographical locations.**

I. INTRODUCTION

Cloud computing [1] comes with several inherent capabilities such as scalability, on-demand resource allocation, and easy application and services provisioning. It comprises three key service models: Infrastructure-as-a-Service (IaaS), Platform-as-a-Service (PaaS), and Software-as-a-Service (SaaS). However, cloud computing still faces some challenges. The connectivity between the cloud and the end-users is set over the Internet, which may not be suitable for a large set of cloud-based applications such as latency-sensitive Internet of Things (IoT) applications. Well-known examples of such latency-sensitive IoT applications include but are not limited to disaster management, healthcare, smart traffic/accident management, and autonomous driving applications. The IoT, according to the definition considered by a recent survey [2], is "*A global infrastructure for the information society enabling advanced services by interconnecting (physical and virtual) things based on existing and evolving, interoperable information and communication technologies*". To address the limitation of cloud computing, fog computing [3] has been introduced. It is a novel architecture that extends the traditional cloud computing architecture to the edge of the network. This extension results in a hybrid cloud/fog system.

Application providers use PaaS to provision (i.e., develop, deploy, manage, and orchestrate) applications in the cloud. However, existing PaaS solutions (including IoT PaaS solutions) usually focus on cloud computing and do not enable the provisioning of applications with components spanning both cloud and fog, e.g., references [4]-[5]. Provisioning applications that span the cloud and fogs requires novel functions such as application graph generation, which are absent from existing cloud PaaS solutions. These applications are composed of a set of components that interact with different sub-structures such as sequence, parallel, selection, and loop structures [6]. Such applications must be modeled as graphs with these sub-structures, and chains need to be created between the components to define the relationship between them. Furthermore, several functions offered by these existing PaaS systems need to be significantly extended in order to fit in a hybrid cloud/fog environment. This includes, but is not limited to, publication/discovery and migration functions. Fig. 1-b shows a structured graph representation of an IoT application, a smart parade application. The application captures the parade footage and derives visible patterns from the footage. These patterns are analyzed later to identify certain events of interest, such as security threats, ethnicities and the ages of the parade participants. Fig. 2-b shows a structured graph representation for a smart accident management application. This application enables innovative services related to accident management. It decreases the time required for an ambulance to reach the scene of an accident and suppresses the sounds of sirens which can be stressful for the elderly or for infants. With such hybrid cloud/fog systems, some of these IoT applications' components; e.g., latency-sensitive ones, can be hosted and executed in the fog at the edge of the network. These components include the Capture Parade Footage in the smart parade application and the Collision Detection in the smart accident management application. Meanwhile, other components, e.g., those that are delay-tolerant and computationally intensive, can be hosted and executed in the cloud, such as the Historical Storage and the Diagnostics and Prognostics in the smart parade and smart accident management applications, respectively.

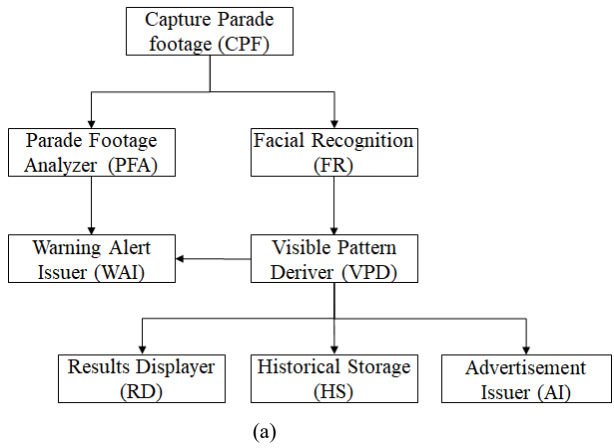
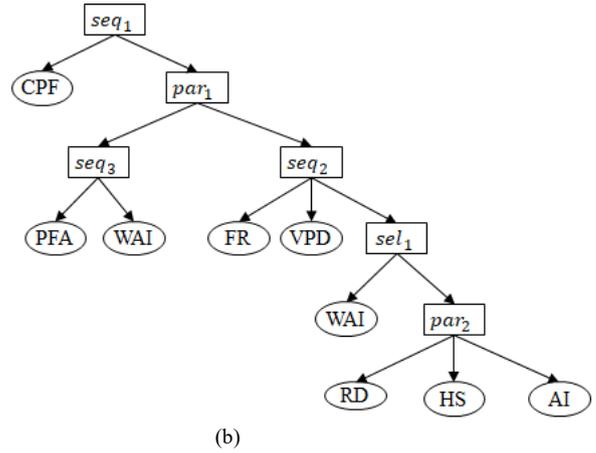

(a)
(b)

Fig. 1 Smart parade application
  (a) Component-based application
  (b) Structured VNF-FG representation

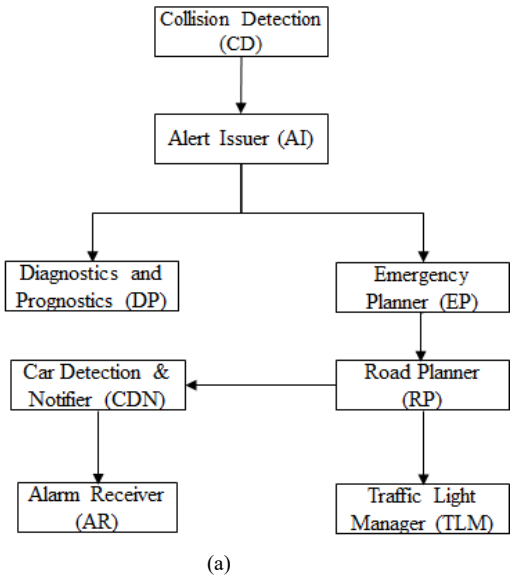
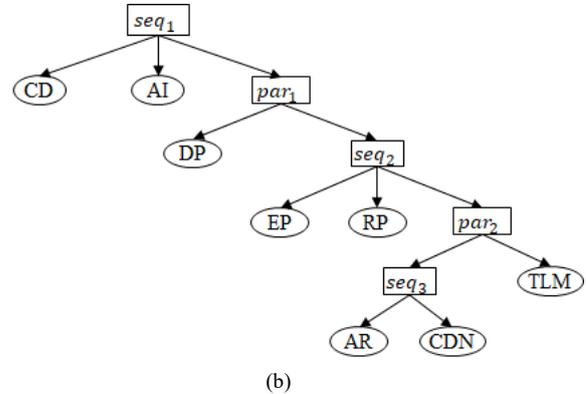

(a)
(b)

Fig. 2 Smart accident management application
  (a) Component-based application
  (b) Structured VNF-FG representation

In this paper, we propose a novel architecture for a Network Functions Virtualization (NFV)-based PaaS for a hybrid cloud/fog system. NFV is an emerging paradigm that employs virtualization as a key technology. Its goal is to decouple network functions from the underlying proprietary hardware and run them as software instances on general-purpose hardware [7][8]. The proposed architecture is IoT use case-driven, and its applications' components are implemented as Virtual Network Functions (VNFs) with execution sequences modeled as graphs. Therefore, the structured graphs representing the applications are VNF Forwarding Graphs (VNF-FG); sets of VNFs chained in a specific order. The proposed PaaS architecture provides full support for the whole provisioning cycle of the application, including development, deployment, management, and orchestration. It automates the provisioning of the applications with components spanning both the cloud and the fog. In addition, it enables the discovery of existing cloud and fog nodes and generates parses application graphs. Moreover, considering a set of interacting components, the proposed architecture enables the creating and updating of chains between application components to keep the application working properly.

The rest of this paper is organized as follows, Section II introduces the two motivating scenarios, describes the challenges and discusses the state-of-the-art. The proposed high-level architecture is presented in Section III, followed by the implementation details, the prototype, and the performance results in Section IV. In the last section, we conclude the paper and outline future work.

## II. STATE OF THE ART

### A. Motivating Scenarios

This section introduces two illustrative motivating scenarios; a smart parade scenario and a smart accident management scenario. These scenarios present in more detail the application graphs depicted in Fig. 1 and Fig. 2. The scenarios highlight the need for an IoT PaaS solution that enables the provisioning of these applications with components spanning both cloud and fog.

*1) Smart Parade Scenario*

We consider a smart parade application to illustrate the motivation behind our work. The application captures parade footage and analyzes it to identify some patterns and/or security threats. The application can be composed of several components, as shown in Fig. 1-a. For instance, the *Capture Parade Footage* component derives visible patterns from the parade footage and sends those patterns to the *Parade Footage Analyzer* for analysis. It can, for instance, identify the clothing brands of most of the people, and send advertisements of those brands more frequently to those people's phones. The application uses *Facial Recognition* techniques to identify the ethnicities and the ages of the various parade participants. This allows advertising companies (through the *Advertisement Issuer* component) to release ads targeting those age groups and ethnicities.

Analyzing the parade footage can also help in identifying security threats. For instance, *Visible Pattern Deriver* can detect any sudden scattering of the crowd, which could be an indication of an altercation/physical fight between a few individuals. Another example is being able to detect if parade participants enter any restricted areas. In such cases, the suspected patterns can be sent to the *Warning Alert Issuer*, where the latter notifies the respective authorities (Ambulance, police, etc.). In addition, all the derived patterns can be sent to a *Historical Storage* system for long term storage and to a *Results Displayer* component to display results relevant to the parade (such as the total number of participants).

*2) Smart Accident Management Scenario*

Smart transportation is an important pillar for the quality of life of citizens in a city. According to the World Health Organization (WHO) 2013, the total number of road traffic deaths is 1.24 million per year worldwide, while the number of injuries caused by crashes is more than 20 million [9]. Accordingly, we consider a smart accident management application that offers innovative services related to accident management. This application decreases the time needed for an ambulance to reach the scene of an accident and omits the sounds of sirens, which can be stressful for the elderly and for infants.

This application can be composed of several components, as shown in Fig. 2-a. For instance, a *Collision Detector* can detect collisions/crashes and share the location of the crash to an *Alert Issuer* on the nearest Road Side Unit (RSU). The *Alert Issuer* informs the *Emergency Planner* for real-time emergency response management.

The application can also find the shortest path between the accident scene and the emergency vehicle through a *Road Planner* component. This component shares the real-time location of the ambulance with a *Car Detector & Notifier* component, which is originally hosted on the RSU closest to the ambulance's initial location. The *Car Detector & Notifier* keeps migrating to RSUs one step ahead of the ambulance in order to detect all the cars on the same street and direction as the ambulance. It sends a message to cars to move to the right so that the ambulance can move easily. The *Car Detector & Notifier* can also coordinate with a *Traffic Light Manager* component to facilitate and accelerate the movement of the ambulance. In addition, all the accident data can be sent to a *Diagnostics & Prognostics* component for further analysis and long-term storage.

### B. Challenges

The identified challenges cover the whole IoT application's lifecycle, i.e., development, deployment, execution, management, and orchestration.

*1) Development Phase Challenges*

Developing IoT application components that can be hosted and executed in either a cloud or a fog is one of the major challenges in the application development phase. In the smart parade application, the *Visible Pattern Deriver* is latency-sensitive and so it may be better to host it in the fog, while the *Historical Storage* component is delay-tolerant and thus can be hosted in the cloud. Similarly, in the smart accident management application, the *Collision Detector* can be hosted in the fog, while the *Diagnostics & Prognostics* component can be hosted in the cloud.

Generating application graphs pose another challenge. The main reason is that the application is composed of a set of interacting components that can be executed in sequence, parallel, selection, and loop, as in the smart parade scenario. Accordingly, they need to be modeled as graphs with these substructures. Specifying the applications' QoS requirements, such as the deadline threshold, is another challenge.

TABLE I. SUMMARY OF THE RELATED WORKS

| Scope | Papers | Challenges | | | | | | | | | |
|---|---|---|---|---|---|---|---|---|---|---|---|
| | | Development | | | Deployment | | Management | | | Orchestration | |
| | | Host apps in cloud/fog | App graphs | Specify QoS | Discover cloud/fog nodes | Deployment plan | Control interface | Migration plan | Create chains | Orchestrate cloud/fog | Parse graphs |
| Hybrid Cloud/Fog | Yangui et al. [4] | ✓ | x | ✓ | x | x | ✓ | ✓ | x | ✓ | x |
| | Pahl et al. [10] | ✓ | x | x | x | x | x | x | x | ✓ | x |
| | Liyanage et al. [5] | ✓ | x | ✓ | ✓ | ✓ | x | x | x | x | x |
| Fog Systems | Yigitoglu et al. [11] | ✓ | x | ✓ | x | ✓ | x | x | x | ✓ | x |
| | Saurez et al. [12] | ✓ | x | ✓ | ✓ | x | ✓ | ✓ | x | ✓ | x |
| | Tao et al. [14] | x | x | x | ✓ | x | ✓ | x | x | x | x |
| | Tuli et al. [15] | ✓ | x | ✓ | x | ✓ | ✓ | ✓ | x | x | x |
| | Donassolo et al. [13] | x | x | ✓ | x | ✓ | x | ✓ | x | x | x |
| | Liu et al. [16] | ✓ | x | ✓ | x | ✓ | x | x | x | ✓ | x |

*2) Deployment Phase Challenges*

Discovering the cloud and the fog nodes by the PaaS is one of the challenges in the deployment phase. The PaaS should be aware of existing cloud and fog nodes (joining and leaving) with their specifications (e.g., capacity, cost, latency) in order to generate efficient placement plans. Determining such optimal placement plans for application components, given a set of objectives and constraints, is another challenge. For instance, in the smart accident management application, one may envision placing the *Alert Issuer* component in the fog and the *Diagnostics & Prognostics* component in the cloud.

*3) Execution and Management Phase Challenges*

The PaaS needs to interact with both cloud and fog nodes. This is a particular requirement when the PaaS wants to deploy and migrate application components between the cloud and the fog. Accordingly, ensuring there are appropriate control interfaces to enable interoperability at the level of providers and architectural modules is one of the challenges in this phase.

Generating and executing the best migration plans (from cloud to fog and vice versa, also from fog to fog) is another challenge. For instance, in the smart parade scenario, the *Capture Parade Footage* component needs to be migrated between fog nodes along with the parade movement. Similarly, in the smart accident management scenario, the *Car Detector & Notifier* component needs to be migrated between fog nodes (i.e., RSUs) one step in advance of the ambulance to clear the way for the ambulance to pass swiftly.

Creating and updating chains between the components is another challenge. For instance, if a component is migrated to another fog node, there is a need to update the chain to keep the application working properly.

*4) Orchestration Phase Challenges*

The first challenge in this phase is to have an orchestrator in the PaaS for coordination purposes. This is required in order to orchestrate the cloud/fog resources and manage the application's lifecycle including deployment, chaining, execution, monitoring, and migration. In addition, the orchestrator needs to execute different orchestration plans such as deployment plans, migration plans, etc. Yet another challenge is to parse the application graph and derive the chaining plan.

*C. The State-of-the-Art and its Shortcomings*

In this section, we review the relevant literature on architectures for hybrid cloud/fog systems. In the first subsection, we review the proposed PaaS architectures for hybrid cloud/fog environments. We then review the proposed architectures for fog systems where the proposed architectures are either fog architectures or architectures spread over the fog and the cloud. Table I provides a summary of the papers reviewed in this section, in which we outline the challenges addressed by each paper.

*1) Architectures for PaaS for Hybrid Cloud/Fog Systems*

Relatively few works have proposed PaaS architecture for hybrid cloud/fog systems. Yangui et al. [4] propose a PaaS architecture for a hybrid cloud/fog system composed of four layers: development, deployment, hosting and execution, and management. Their proposed architecture is able to specify the applications' QoS requirements using the SLA Manager module. It also has appropriate control interfaces to enable interoperability between the PaaS and the fog. For the development phase, their proposed architecture provides an Integrated Development Environment (IDE) to enable the development of application components that can be hosted on either the cloud or the fog. However, this IDE uses existing application development frameworks to provide developers with the tools facilitating such development. Moreover, the proposed PaaS does not enable the discovery of newly joining or leaving cloud and fog nodes. The existing cloud/fog nodes are pre-configured. It also does not enable the optimal placement plan for these components to be determined. Pahl *et al.* [10] present a container-based edge cloud PaaS architecture. Their proposed architecture enables fog nodes to run their applications in containers as well as the orchestration of the deployment of those containers. While their proposed architecture includes a development layer to provision and manage applications over cloud/fog nodes, it does not support the discovery of cloud/fog nodes or the generation of the best deployment plan. The proposed architecture enables the migrating of containers, but it does not enable the best migration plans to be generated.

In contrast to Yangui *et al.* [4] and Pahl *et al.* [10], the PaaS architecture proposed by Liyanage *et al.* [5] enables generating the best deployment plan by proposing a component distribution scheme. In addition, they incorporate a publication/discovery mechanism for the underlying node's specifications using Service Description Metadata (SDM). Their main contribution is proposing a service-oriented PaaS architecture that allows users to deploy and execute their own applications on cloud and mist resources. Mist was proposed to reduce the burden on the fog. The proposed architecture supports resource-aware autonomous service configuration and takes the QoS requirements of an application into consideration. Although the publication mechanism is based on RESTful services, the interfaces between the remaining architectural modules are not discussed. In addition, none of the remaining challenges discussed in our paper are addressed by this proposed architecture.

*2) Architectures for Fog Systems*

Several works have proposed architectures for fog systems. Most of these architectures are designed to span the cloud and the fog, such as the architectures proposed by Yigitoglu *et al.* [11] and Saurez *et al.* [12]; only one architecture is strictly fog architecture, the architecture proposed by Donassolo *et al.* [13]. Yigitoglu *et al.* [11] propose a framework called Foggy that facilitates dynamic resource provisioning and automates application deployment in fog computing architectures. Foggy assumes that IoT devices can host Docker containers. It has three-tier architecture: edge devices (e.g., fog nodes), a network infrastructure to connect the edge devices to the cloud, and cloud services. The focus of the proposed framework is on the deployment and the orchestration phases. For example, it enables determining an optimal deployment plan for an application. However, it does not enable the fog nodes to be discovered dynamically and instead assumes a pre-configured list of the nodes. Moreover, creating chains between different application components and migrating application components among different nodes are not discussed. In the development phase, the developers push their containerized application packages and their specifications to the orchestrator to ensure the QoS for each application. The orchestrator is a central entity and is in charge of monitoring the nodes' resources.

Saurez *et al.* [12] propose a framework called Foglet that facilitates distributed programming across the resources from IoT devices to the cloud. Their proposed framework provides communication APIs for discovering fog resources. It also enables QoS-aware incremental deployment over different fog nodes via containerization. Foglet first places application components at the lowest layer, and gradually finds the best candidates in upper layers; hence it does not enable any determining of the optimal deployment plan. In the proposed Foglet framework, fog provides interfaces that allow its computing instances to be managed. Migrating application components among fog nodes is also supported. The orchestrator is responsible for the deployment and migration of the application components. However, it is not capable of parsing application graphs. In addition, there is no discussion on how to create or update chains among different application components. Tao *et al.* [14] propose an architecture called Foud that can facilitate the growth of Vehicle to Grid (V2G) services and applications. Their proposed architecture is organized over three layers: the user layer, which is composed of different types of end-users in V2G systems, the service layer, which is divided into two sub-layers: cloud and fog, and the network layer. The network layer provides an interconnection between the cloud and the fog. It basically provides protocol, interface, and security techniques. Accordingly, interoperability between the two sub-models is achieved. However, most of the execution and management layer challenges are not discussed, such as migrating applications/components between cloud/fog nodes (which is critical to support the mobility of end-users and fog nodes), and chaining application components. In addition, orchestrating the cloud/fog resources and managing applications' lifecycles are not discussed in this proposed architecture. Finally, the proposed architecture does not ensure the applications' desired level of QoS.

Tuli *et al.* [15] propose a lightweight framework called FogBus to integrate IoT, fog, and cloud infrastructures. Their proposed framework uses blockchain mechanisms to provide secure and authenticated data transfer between IoT devices, fog nodes, and cloud data centers. It also enables implementing resource management and scheduling policies for applications spanning the cloud and the fog. The proposed FogBus framework can generate optimal deployment plans using the resource manager module. This module identifies the requirements of different applications and selects the suitable resources to execute the applications accordingly, thereby determining optimal application placement plans. This framework is also capable of monitoring the applications to ensure the QoS requirements are met. In the case of QoS violation, the framework initiates application migration. However, creating and updating chains between the components is not discussed. The applications' details are maintained in a catalog that contains information about different applications, including their operations, resource requirements, and dependencies. However, it is not mentioned if this catalog supports application graphs with interacting components using different substructures. The proposed FogBus framework has REST-based interfaces to exchange data and share information among different nodes. Hence, it enables interoperability. Finally, the proposed framework does not enable the discovery of the underlying joining and leaving fog nodes.

Donassolo *et al*. [13] propose an orchestration framework for the automation of the deployment, the scalability management, and the migration of component-based IoT applications. Their proposed solution offers a general centralized framework for holistic fog resource orchestration and application orchestration. Using this framework, application components can be deployed on either the end-devices or the fog nodes. It is not discussed if the components can be deployed over the cloud nodes. The framework also includes a module called a service descriptor that describes the application, its components, and the components' requirements. However, it is not clear if this module can describe application graphs with interacting components using substructures such as selection and parallel. The service manager module of the proposed framework can deal with dynamic applications and trigger migration actions when necessary. However, generating and updating chains between the components is not discussed. In addition, having appropriate control interfaces to enable interoperability is not supported. The proposed framework proposes a strategy to determine the optimal placement plans of IoT-based application components such that a guaranteed QoS can be ensured. However, it is not capable of discovering the underlying fog nodes (joining/leaving) when generating the placement plan.

Liu *et al*. [16] propose a fog computing architecture for resource allocation. It considers latency reduction combined with reliability, fault tolerance, and privacy. This fog computing architecture is elaborated in two parts: computing and networking. Four layers are considered for the computing part: a hardware platform, a software and virtualization platform, functional components, and a fog computing applications interface. The networking side is composed of three layers: wireless technology, single-hop/ad-hoc communications, and a software-defined network concept. The authors formulate the resource optimization problem considering the QoS in terms of latency and use a genetic algorithm to solve it. Hence, this approach supports generating the best deployment plan. In addition, they consider both the fog and the cloud for hosting application components. The proposed architecture includes an orchestration that is responsible for analyzing, planning, and executing a task. However, none of the remaining challenges presented in our paper is addressed by the proposed fog computing architecture.

It should be noted that none of the works presented here enable generating or parsing application graphs to model the interactions between different components of an application. In contrast, we introduce a novel module that can generate application graphs as well as model the interactions between the different application components. In addition, none of the presented papers enable the creating or updating of chains between the application components. These chains are necessary when, for instance, a component is migrated from one node to another, where the chain needs to be updated such that the application works properly. Moreover, several functions offered by existing PaaS need to be significantly extended in order to fit in a hybrid cloud/fog system, such as the publication/discovery function proposed in [5] [12] [14].

### III. PROPOSED IoT PaaS ARCHITECTURE FOR NFV-BASED HYBRID CLOUD/FOG SYSTEMS

This section presents a high-level architecture of the proposed IoT PaaS for hybrid cloud/fog systems. An overview of the designed architecture is first introduced, followed by a discussion of the architectural modules and the interfaces. This section ends with the presentation of an illustrative sequence diagram.

#### A. Architecture Overview

A high-level view of the proposed architecture is depicted in Fig. 3. It includes the PaaS, the Cloud Domain(s), and the Fog Domain(s). It should be noted that the PaaS could be running in the Cloud Domain, in the Fog Domain, or be provided by a third party. It can also be distributed across several domains. For instance, if we take the smart parade application, cameras could be distributed along the roads of the parade route to capture parade footage. Accordingly, some of the application components (e.g., Capture Parade Footage) will be distributed to improve its effectiveness. In such cases, it is better to distribute the PaaS across several domains to ease the development, the deployment, the management, and the orchestration of the application. The IoT PaaS is distributed over four layers: An Application Development layer, an Application Deployment layer, an Application Execution and Management layer, and an Application Orchestration layer.

We present the modules in each layer of the PaaS and the modules in the cloud and fog domains below. This is followed by a presentation of the interaction interfaces between the different modules and a description of the main procedures. It should be noted that some of the modules of the proposed architecture are novel, such as the *App. Graph Generator* and the *Infrastructure Repository*. These modules are depicted in yellow in Fig. 3. Some other modules are extended modules from traditional PaaS architecture, shown in blue.

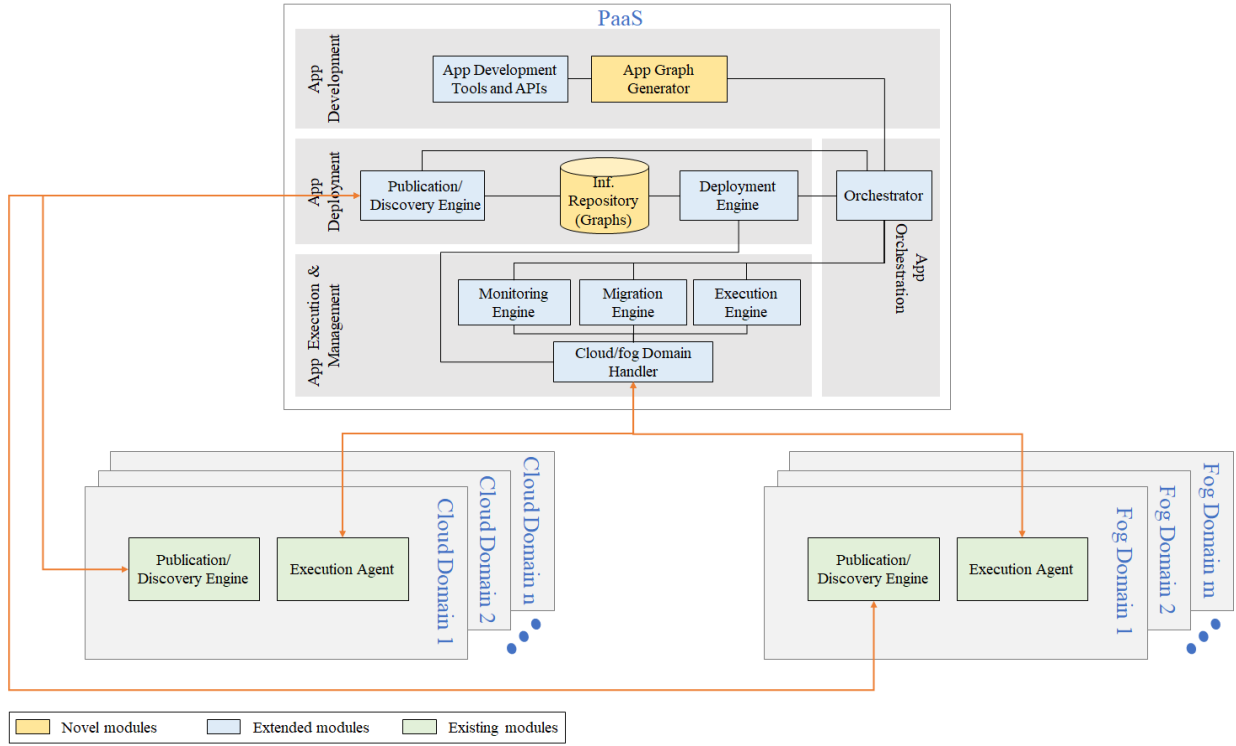

Fig. 3. High-level architecture of IoT PaaS for hybrid cloud/fog system

1) Architectural Modules
   a) Modules in the PaaS

**Application Development Layer**

This layer contains two modules: The *App Development Tools and APIs* and the *App Graph Generator*. The *App. Development Tools and APIs* module includes different tools and APIs to give developers an environment for developing IoT applications. It is an extended module of traditional PaaS, as it provides developers with tools to facilitate the development of applications that span both the cloud and the fog. This module can provide standard/industrial development tools and APIs such as Eclipse, as well as proprietary development tools and APIs. For instance, Google Compute Engine binds the developer to a specific platform offered by the vendor. An application developed using the Google API can only run on a particular environment, and so the possibility of extensibility beyond a specific vendor's support is quite limited. In contrast, Cloud Foundry supports applications developed in any of the standard development tools. The *App. Graph Generator* is a novel module. It generates a graph for an application and a description of each component. The graph models the interaction between different application components. For instance, in the case of a scenario presented in Section II; the smart accident management application, the *App. Graph Generator* will generate as output a graph as shown in Fig. 2-b.

**Application Deployment Layer**

This layer includes the *Infrastructure Repository*, the *Deployment Engine*, and the *Publication/Discovery Engine*. The *Infrastructure Repository* is a novel module that allows the storage of graph-like data. It uses a graph structure with nodes, edges, and properties to represent and store data. This data includes information about the cloud and the fog nodes, such as their capacity and relationships.

TABLE II. EXAMPLES OF SOME OF THE API OPERATIONS EXPOSED BY THE PUBLICATION/DISCOVERY ENGINE TO THE ORCHESTRATOR

| REST Resource | Operation | HTTP Action and Resource URI |
|---|---|---|
| List of Domains | Get list of domains | GET:/domains |
| List of fog nodes | SUBSCRIBE to the information of a list of fog nodes | POST:/fognodes?fromuri={subscriberuri} |
| List of fog nodes | Unsubscribe from the information of a list of fog nodes | DELETE:/fognodes?fromuri={subscriberuri} |

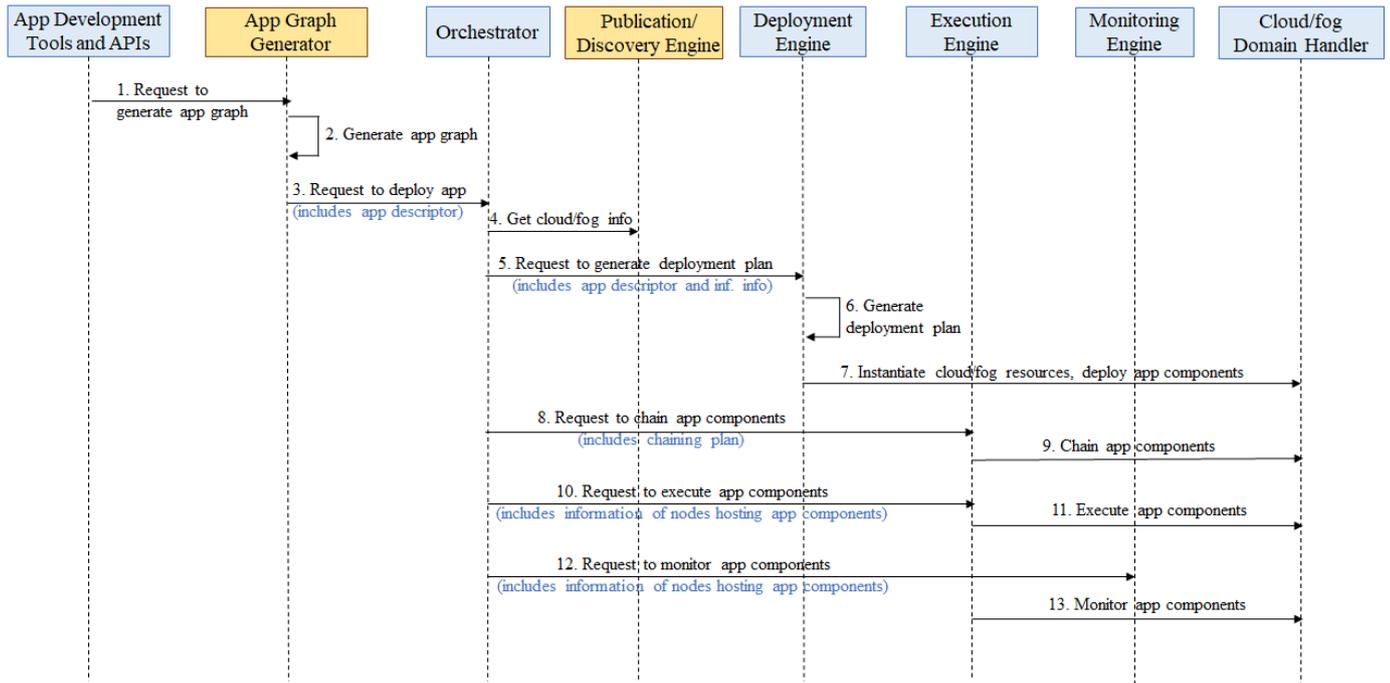

Fig. 4. Sequence Diagram for the Orchestrator Deployment Plan (Application deployment procedure)

The *Deployment Engine* is an extended module of regular PaaS in terms of considering the fog infrastructure. It is responsible for finding the optimal deployment plan of IoT application components over the cloud and fog infrastructures. To that end, it runs a placement algorithm, such as the one presented in [4]. For instance, let us consider both applications presented in Section II, the smart parade application and the smart accident management application. They consist of a set of interacting components that represent a VNF-FG. The placement algorithm presented in [4] finds the near-optimal placement of this VNF-FG over the cloud and fog infrastructures (i.e., NFVI) such that the application execution time and cost are minimized. The *Deployment Engine* instantiates the cloud/fog resources required for hosting and executing the applications' components (e.g., service containers) and processes the deployment of the application's components over these resources. The *Publication/Discovery Engine* another extended module, is responsible for the publication and discovery functions that locate the cloud nodes/resources as well as the fog nodes/resources. Accordingly, it constructs a graph structure representing the relations among the cloud and the fog nodes.

**Application Execution and Management Layer**

Four modules are included in this layer: The *Monitoring Engine*, the *Migration Engine*, the *Execution Engine*, and the *Cloud/Fog Domain Handler*. It should be noted that all the modules in this layer are extended modules from a traditional PaaS in terms of handling the fog infrastructure. The *Monitoring Engine* monitors the cloud/fog resources to detect mobility, bottlenecks, etc. The *Migration Engine* runs a migration algorithm, similar to the one presented in [17]. Considering the smart parade application, when the Capture Parade Footage component needs to be migrated between the fog nodes, the algorithm finds the best node to migrate to, and in an acceptable time. The *Migration Engine* also performs the actual migration of application components. The *Execution Engine* is responsible for creating or updating chains between application components as well as for executing the application components. The *Cloud/Fog Domain Handler* is an extension of the IaaS communication component in conventional PaaS architectures. It handles all the communications between the PaaS and the cloud and fog infrastructures.

**Application Orchestration Layer**

This layer includes the *Orchestrator* which is in charge of orchestrating the cloud/fog resources. It is also an extended module of traditional PaaS architecture. It is responsible for managing the lifecycle of the application, including deployment, chaining, execution, monitoring, and migration. It can execute different orchestration plans according to the requests it receives, such as a Deployment Orchestration Plan and a Migration Orchestration Plan.

*b)* *Modules in the Cloud/Fog Domains*

The *Publication/Discovery Engine* is responsible for the publication and discovery function of the nodes in its domain. The *Execution Engine* provides the necessary execution environment (e.g., containers) for the cloud and fog nodes to execute the application components.

*2) Interfaces*

The general principle for designing the interactions between the different modules and the different domains is the use of the REpresentational State Transfer (REST) architectural style. All of the interfaces expose CRUD (i.e., Create, Read, Update, and Delete) operations. Table II gives some examples of the proposed REST interface for the interactions between the *Orchestrator* and the *Publication/Discovery Engine* modules. This interface defines the resources on the *Publication/Discovery Engine* and allows the *Orchestrator* to (un)subscribe to the information of a list of fog nodes hosting application components. It also allows the *Orchestrator* to get the list of cloud/fog domains along with their cloud/fog nodes.

*3) Procedures*

The proposed architecture includes the following procedures: application development, application deployment, and application migration. We describe the application deployment and migration procedures below.

*a) Application Deployment*

The process is initiated when the *Orchestrator* receives a request from the Application Development layer to deploy an application. This request includes the IoT application descriptor (i.e., the application graph and the descriptor of each component). The *Orchestrator*, as part of the Deployment Orchestration Plan, first gets the cloud/fog infrastructure information from the *Infrastructure Repository*. It then sends the infrastructure information along with the application descriptor to the *Deployment Engine*. The latter runs a placement algorithm to generate a deployment plan. According to the deployment plan, the *Deployment Engine* instantiates the cloud/fog resources required for hosting and executing the application's components (e.g., service containers) and processes the deployment of the application's components over these resources. The *Orchestrator* then asks the *Execution Engine* to generate a chaining plan. The latter chains the application components according to the chaining plan and begins executing the components. Once the execution of the application is initiated, the *Monitoring Engine* starts monitoring the application components.

It should be noted that the proposed IoT PaaS architecture supports the on-demand discovery of the cloud/fog resources. This process is initiated when the *Orchestrator* receives a request from the App Development layer to deploy an application. In response, the *Orchestrator* asks the *Publication/Discovery Engine* to discover the cloud/fog resources and then gives that information to the *Deployment Engine* along with the application descriptor to generate a deployment plan.

*b) Application Migration*

This process is initiated when the *Orchestrator* receives a request from the *Monitoring Engine*. The *Orchestrator*, as part of the Migration Orchestration Plan, first processes the request and decides which component needs to be migrated. It then sends a request to the *Migration Engine* to generate the best migration plan. The *Migration Engine* runs a migration algorithm [17] and finds the best node to migrate the application component. Once the component has migrated, the *Orchestrator* sends a request that includes the new node hosting the application component to the *Monitoring Engine* so that it can monitor all the application components.

*B. Illustrative Sequence Diagram*

Fig. 4 illustrates a sequence diagram of the interactions of different architectural modules during the application deployment phase. It is assumed that an initial discovery of cloud/fog nodes has already been done and that their information is in the *Infrastructure Repository*. During the deployment, the *App Graph Generator* sends the IoT application descriptor to the *Orchestrator* and requests it to deploy the application (Fig. 4, actions 1, 2, and 3). The *Orchestrator* then gets the information about the cloud and fog nodes and sends it to the *Deployment Engine* to generate a deployment plan (actions 4 and 5). The *Deployment Engine* generates a deployment plan, instantiates the cloud and fog resources, and performs the actual deployment of the application components (actions 6 and 7). Once the components are deployed, the *Orchestrator* sends a request to the *Execution Engine* to generate a chaining plan. The latter then chains the application components (actions 8 and 9). Execution of the application components is then started by the *Execution Engine* (actions 10 and 11). Finally, the *Orchestrator* sends a request to the *Monitoring Engine* to monitor the application components (actions 12 and 13).

IV. IMPLEMENTATION AND EXPERIMENTATIONS

*A. Implementation Scenario*

The smart parade application presented in Section II.A-1 was implemented in a prototype. The application captures parade footage and sends it for analytics, in which facial recognition techniques are utilized to identify and display each person's ID, age, and gender. In the fog domains, only the information received from the cameras in the same fog domain is displayed. However, the footage received from all the fog domains is displayed in the cloud. In other words, the cloud acts as a centralized displayer for the information displayed at each fog domain. The reader should note that the identification of gender and ages could trigger several value-added services, as explained in section II.A-1. It should also be noted that, as the parade moves, the application migrates the application components residing in the fog, namely the machine learning module that is responsible for facial recognition analysis of the people captured in the video footage, and the results displayer that displays the results of the machine learning module.

Accordingly, the following application components are implemented as VNFs:

(1) *Capture Parade Footage* - where the camera manager resides; it starts/stops/pulls out footage from the camera;

(2) *Parade Footage Analyzer* - includes a machine learning (ML) module that can determine the ID, gender, and the age of the participants; and

(3) *Results Displayer* - displays the ID, age, and gender for each face captured by the cameras.

Two IP cameras are used to capture the parade footage. One is an Axis M1031 network camera[2] with IEEE 802.11b and IEEE 802.11g network interface, and the other is an Axis M1065 LW Network Camera[3] with IEEE 802.11b, IEEE 802.11g, and IEEE 802.11n network interface. Both support wired and wireless communication and each contain a microphone and a speaker.

B. *Prototype Implementation*

Fig. 5 shows the PaaS prototype architecture. The software architecture of Cloudify is reused for our PaaS implementation. As shown in Fig. 5, the *Application Graph Generator*, the *Publication/Discovery Engine*, the *Deployment Engine*, the *Orchestrator*, and the *Migration Engine* are implemented, but the *Monitoring Engine* and the *Execution Engine* modules are not implemented.

The parade scenario presented in Section IV.A is implemented in this prototype using the *Parade Footage Analyzer* component and the *Results Displayer* component.

Cloudify[4] is an open-source cloud orchestration framework that enables modeling applications and services and automates their entire life cycle. An application in Cloudify is described in a blueprint and its DSL (Domain Specific Language) is based on the TOSCA standard. The blueprints are YAML documents and are used to describe how the application should be deployed, managed, and automated.

Nodes can be defined in the blueprints. These nodes represent the services. Each node has its own properties and some unique features. In this prototype, we define the following nodes in the blueprints: the graph generator node, the deployment node, the publication/discovery node, the orchestrator node, and the migration node. Accordingly, using the blueprints, Cloudify orchestrates the execution of the *App. Graph Generator*, the *Deployment Engine*, the *Publication/Discovery Engine*, and the *Migration Engine*. These nodes act as REST clients using the Cloudify REST plugin in order to communicate with different architectural modules and nodes.

The *Application Graph Generator* is implemented using Java Swing libraries. We implemented it as a simple Java desktop application that generates description files based on the user input. This input contains various information about the application and the relationship between its components, including information about performance requirements of the application (e.g., required traffic, memory size, disk size, etc.)

For the *Publication/Discovery Engine*, a publication node in a Cloudify blueprint acts as a REST client using the Cloudify REST plugin. It sends a request to the *Publication/Discovery Engine* in each fog domain in order to get the most updated fog nodes' information. It then stores this information in a runtime property inside the Cloudify framework. In the prototype, we assumed that we have one cloud node, hence no need to discover it.

For the *Orchestrator*, the orchestrator node (a Cloudify blueprint) uses the Cloudify REST plugin to communicate with different architectural modules and to cooperate among them. It receives the application description from the *App Graph Generator*. This information includes the interaction between different components and the description of each component. It also uses the same plugin to receive the underlying cloud/fog nodes' information from the *Publication/Discovery Engine*. It then merges this information in the blueprint, and, using the REST plugin of Cloudify, sends this information to the *Deployment Engine*.

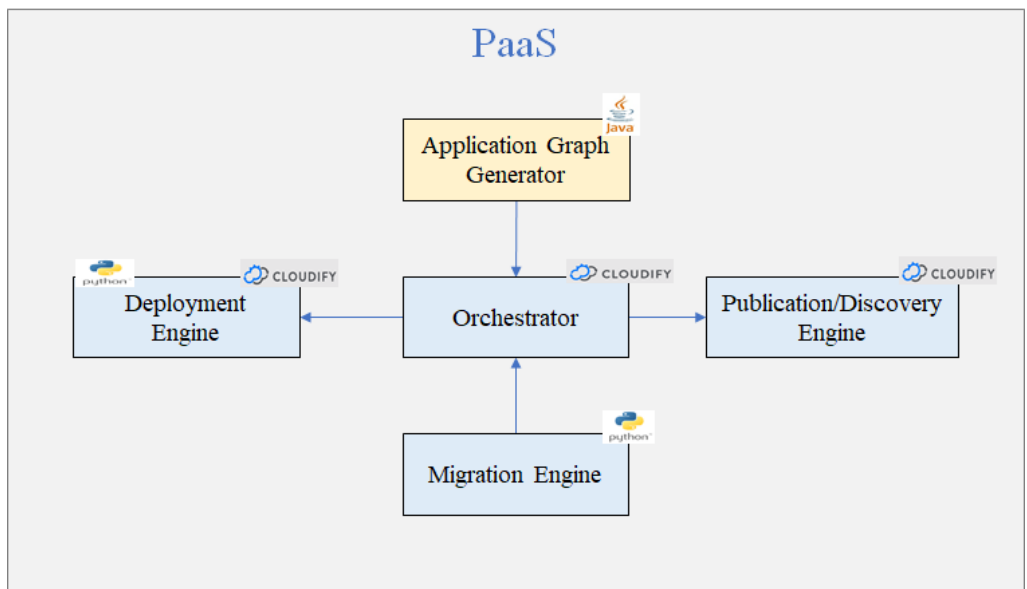

Fig. 5. The prototype architecture

---

[2] /axis.com/en-ca/products/axis-m1031-w
[3] /axis.com/en-ca/products/axis-m1065-lw
[4] /cloudify.co/

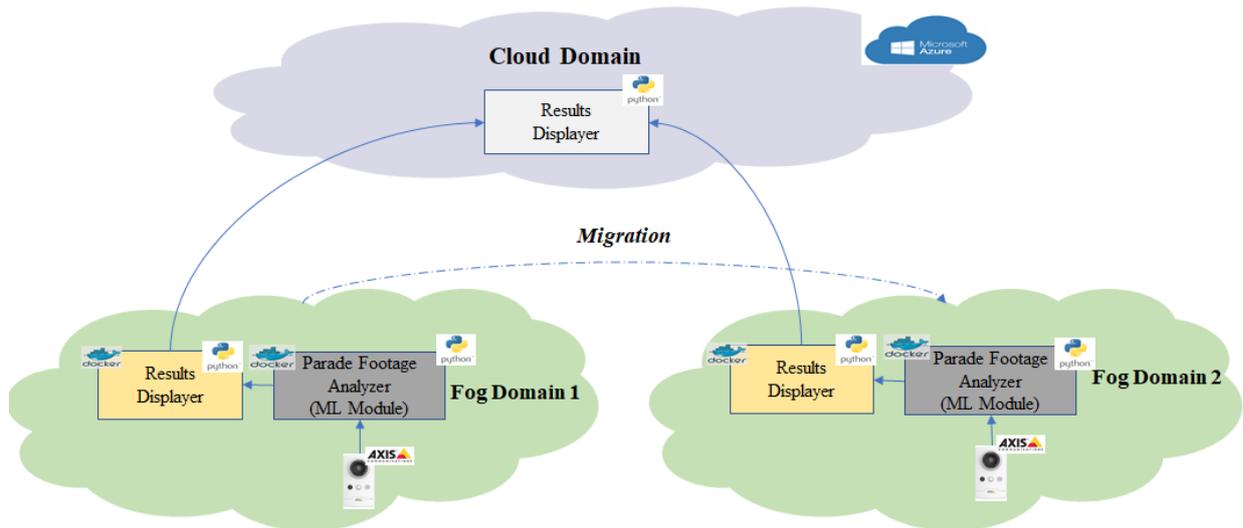

Fig. 6. A prototype view of the implementation scenario

The *Deployment Engine* is implemented as a python-based web application using the python web framework Flask. The deployment process relies on a couple of Docker containers to launch both the results displayer and the Machine Learning module on the target fog node. This node, described in a Cloudify blueprint, uses the Cloudify REST plugin to receive data (i.e., the application descriptor and the cloud/fog nodes information) from the *Orchestrator*. It then sends a request to the *Deployment Engine* to deploy the application components (implemented as VNFs) on the cloud/fog nodes. This blueprint uses the Cloudify Fabric plugin to communicate with the *Deployment Engine*. The Fabric plugin enables Cloudify to SSH into the respective fog node in order to deploy the application component on it. In addition, this blueprint contains additional details about the nodes and the scripts needed during the deployment process.

For the *Migration Engine*, the migration node in the Cloudify blueprint sends a request to the *Migration Engine* using the Cloudify REST plugin to start migrating the *Capture Parade Footage* and the *Results Displayer* components from one fog node to another. The *Migration Engine* is implemented using the python web framework Flask, relying on Docker containers to migrate both components residing on fog nodes (i.e., *Results Displayer* and *Parade Footage Analyzer*) from one fog node to another.

Fig. 6 demonstrates a prototype view of the implementation scenario. The application components are implemented as VNFs. The VNFs are packaged in Docker containers and are pushed to the DockerHub repository. Whenever a VNF needs to be migrated from one fog node to another, the *Migration Engine* sends a request to the first fog node to stop the container. The fog node then pushes the container image to the DockerHub repository, from which the second fog node pulls the container image and runs the container.

For the *Parade Footage Analyzer* (ML Module) component, we used a python application that can directly access the IP camera by specifying the camera's URL and thus obtains real-time video streams[5][6]. This ML application recognizes the age and the gender of the people in front of the camera and tags each face with the detected age and gender. The photo is taken from the live camera stream by the cv2 module (a python library designed to solve computer vision problems), which then converts the image to grayscale to detect faces. The cropped faces are used later to feed the neural network model for prediction purposes. These results are then sent from the *Parade Footage Analyzer* to the *Results Displayer* via a REST API (Flask-REST app). Flask is a lightweight WSGI web application framework, and Flask-REST is an extension for Flask that adds support for building REST APIs. The *Results Displayer* component is implemented using Flask. It exposes a REST API implemented as a Flask web app to the *Results Displayer* (on the cloud) and to the *Parade Footage Analyzer*.

---

[5] /data.vision.ee.ethz.ch/cvl/rrothe/imdb-wiki/

[6] /lology.com/blog/easy-real-time-gender-age-prediction-from-webcam-video-with-keras/

*C. Setup*

The PaaS runs on a machine with dual 2X8-Core 2.50GHz Intel Xeon CPU E5-2450v2 and 40GB of memory in one setting and it is distributed between the local machine and Microsoft Azure cloud in another setting. In a distributed PaaS setting, the machines used (in Virginia and Iowa) in Microsoft Azure have 4Go of RAM with 2 vCPUs Intel® Xeon® CPU E5-2660 0 @ 2.20GHz and Ubuntu Server 18.04. The prototype includes one cloud node and two fog nodes. The cloud node is a Virtual Machine (VM) on the Microsoft Azure cloud. The VM has an Intel® Xeon® CPU E5-2660 0 @ 2.20GHz (2 CPUs) with Windows 10 Pro 64-bit. The first fog node (i.e., fog node 1) is a laptop with an Intel® Core i7-2620M 2.70GHZ CPU with 8GB of RAM running Ubuntu 18.04.2, and the second (i.e., fog node 2) is another laptop with an Intel® Core i5-2540M 2.60GHz CPU with 4GB of RAM running Ubuntu 18.04.2.

*D. Performance Evaluations*

*1) Performance Metrics*

**Orchestration Latency** - measured from the time a request to deploy an application to the orchestrator is initiated to the time the acknowledgment of orchestration is received. Orchestration latency is measured for executing both the deployment plan and the migration plan. The deployment plan includes the discovery of cloud/fog nodes, application deployment, chaining, execution, and monitoring. The migration plan includes application migration and monitoring. In addition, the orchestration latencies for centralized and distributed PaaS are also calculated considering different distributions of the PaaS modules. For executing the deployment plan and the migration plan, different test cases have been considered (i.e., test cases 4, 5, and 6).

**End to End (E2E) delay** – measured from the time the cameras send footage to the time the cloud R*esults* Displayer displays the final results. We vary the placement of the components and show the effect of changing the placement.

*2) Test Cases*

The first three test cases consider the PaaS as a centralized entity, where all its modules are deployed on a local machine in our lab in Montreal. However, they consider different distribution of application components. The remaining test cases consider a distributed PaaS with a different distribution of its modules (mainly the Deployment Engine and the Migration Engine). However, they consider application components running on the same node.

**Test Case 1** – This test case considers an environment composed of two fog nodes and one cloud node. Similar to the description of the prototype architecture, the *Parade Footage Analyzer (ML Module)* and the fog's *Results Displayer* are each deployed on a fog node (i.e., a laptop), while the cloud *Results Displayer* is deployed in the cloud.

**Test Case 2** – This test case considers an environment with only two fog nodes. All the components are deployed on the fog nodes. The first fog node runs the fog *Results Displayer* while the second fog node runs the cloud *Results Displayer* and the *Parade Footage Analyzer*.

**Test Case 3** – This test case considers an environment with one fog node and one cloud node. The *Parade Footage Analyzer* runs on the fog node, while both *Results Displayers* (the one designed for the fog and the one designed for the cloud) run on the cloud.

**Test Case 4** – This test case considers that the Migration Engine and the Deployment Engine are deployed on Microsoft Azure in Virginia while Cloudify and the remaining PaaS modules are deployed on our local machine in Montreal. In addition, it considers all the application components are initially hosted on Microsoft Azure in Iowa and need to be migrated to Microsoft Azure in Virginia.

**Test Case 5** – This test case considers that the Migration Engine and the Deployment Engine are deployed on our local machines in our lab in Montreal, while Cloudify and the remaining PaaS modules are deployed on another machine in our local network in Montreal. Application components are initially running on Microsoft Azure in Iowa and need to be migrated to Microsoft Azure in Virginia,

**Test Case 6** – This last test case considers that the Migration Engine is deployed on Microsoft Azure in Virginia while Cloudify and the remaining PaaS modules are deployed on our local machines in our lab in Montreal. The application components are initially hosted on Microsoft Azure in Iowa and need the be migrated to Microsoft Azure in Virginia.

*3) Results and Discussion*

**Orchestration Latency for Executing the Migration Plan** - Fig. 7 indicates the average latency for executing the migration plan in a centralized PaaS over 15 consecutive experiments conducted for test case 1. We assume that the fog *Results Displayer* and the *Parade Footage Analyzer* are migrated from fog node 1 to fog node 2. The Linux built-in tool time is used again, this time to get the time required to execute the migration plan. The average latency for executing the migration plan is 36.26 sec.

Fig. 8 shows the average latency for executing the migration plan in a distributed PaaS over 15 consecutive experiments for test cases 4, 5, and 6. In test case 4, the Migration Engine is close to the destination node (where we want to migrate the application components) and far from the remaining PaaS modules and the source node hosting the application components. In test case 5, the Migration Engine is closer to the other PaaS modules and far from the source and destination nodes. Finally, in test case 6, the Migration Engine is close to the source node and far from the other PaaS modules and the destination node. The performance results show that the placement of the Migration Engine close to the destination node results in lower latency. Although the difference with the measurements made for the other test cases (Test Cases 5 and

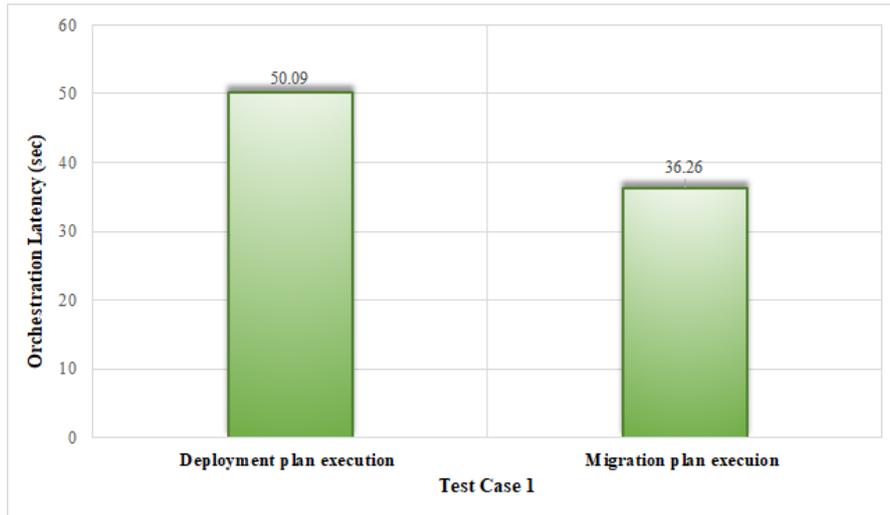

Fig. 7. Orchestration latencies for executing the deployment plan and the migration plan for the parade application considering a centralized PaaS

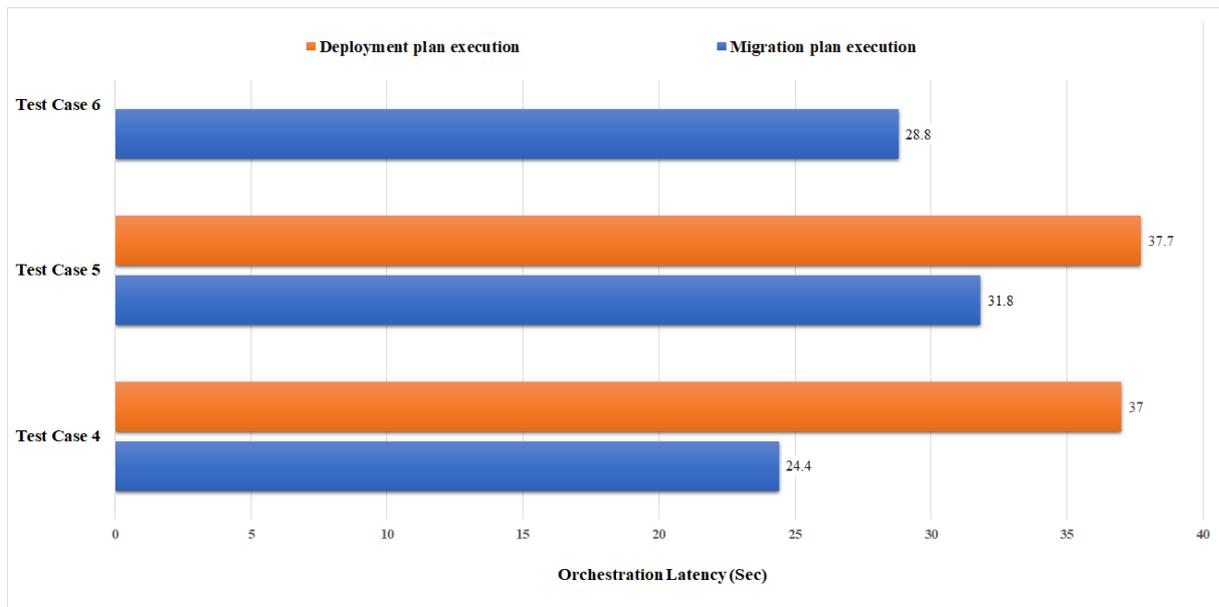

Fig. 8. Orchestration latency for executing the deployment plan and the migration plan for the parade application considering a distributed PaaS

6) was not very big, the PaaS architecture still needs to be combined with a placement algorithm for its modules as well as the application components in order to obtain optimal results in terms of latency.

**Orchestration Latency for Executing the Deployment Plan** - Fig. 7 also indicates the average latency for executing the deployment plan in a centralized PaaS for test case 1, the only test case conducted for this experiment. We used the built-in Linus tool time to get the time required to execute the deployment plan. The results are provided for 15 consecutive experiments. The average latency for executing the deployment plan was 50.09 sec.

Fig. 8 indicates the average latency for executing the deployment plan in a distributed PaaS over 15 consecutive experiments for test cases 4 and 5 only. The same logic for migration was followed for deployment, where the Deployment Engine was first placed closer to the application than the PaaS (test case 4) and then closer to the PaaS modules than the application (test case 5). The results obtained were similar, which shows that the placement of the deployment engine does not influence the execution of the deployment plan for our proposed PaaS architecture.

The procedure for executing the deployment plan involves two additional modules than the procedure for executing the migration plan, hence, the longer average latencies make sense. More specifically, for deployment, the orchestrator has to first communicate with the *Publication/Discovery Engine* and the *App. Graph Generator* before sending a request to the *Deployment Engine* to deploy the application components. However, executing the migration plan only involves sending a request from the O*rchestrator* to the *Migration Engine*, which proceeds to migrate both components (i.e., *Capture Parade Footage* and fog *Results Displayer*) from one fog node to another. It should be noted that for the deployment plan execution, Cloudify orchestrator needs to install two blueprints (one for *Publication/Discovery Engine* and the *App. Graph Generator*, and another one for the *Deployment Engine*). Meanwhile, Cloudify only installs one blueprint for the migration process, since only a single request to the Migration Engine component is needed for this process to take place.

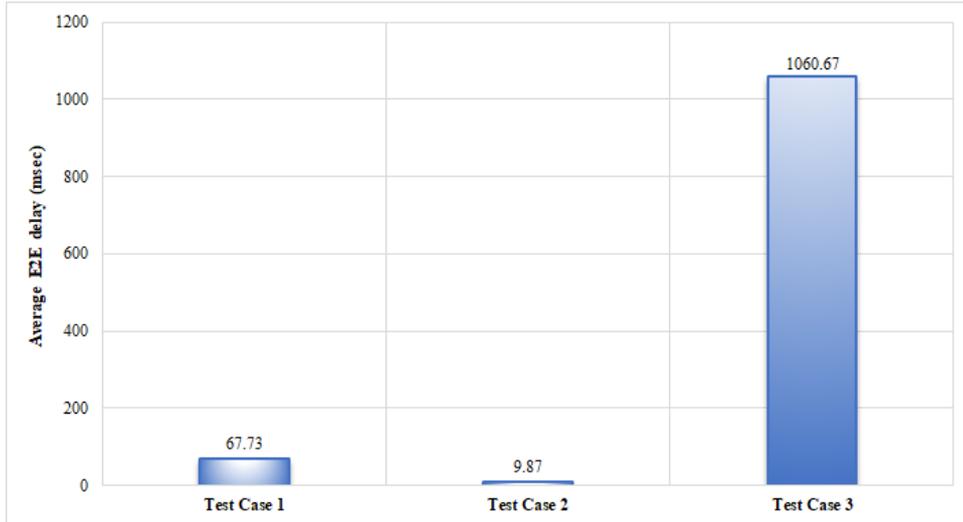

Fig. 9. End to End latency for executing the smart parade application

It should also be noted that the latency for the execution of the migration plan considering a distributed PaaS is lower than the latency in a centralized PaaS. This lower latency is mainly due to the increased networking capabilities of Microsoft Azure compared to the local machines in our lab. The same conclusion can also be made for the difference between the latency for the execution of the deployment plan in a centralized and distributed PaaS.

**End to End Latency** - Fig. 9 shows the end to end latency for executing the implementation scenario presented in Section IV.A. This experiment was conducted over the three test cases (i.e., test case 1, test case 2, and test case 3). The results for 15 consecutive experiments are provided in Fig. 9. The latency is measured via timestamps in the ML module of the *Parade Footage Analyzer* and in the cloud's *Results Displayer* components. The end to end latency can thus be obtained by calculating the time difference between these two timestamps. The lowest latency is obtained in test case 2, where all the components are deployed on the fog nodes. This result is as expected; all the fog nodes are in the same LAN and hence there is very low latency (9.87 msec). Test case 1 shows a relatively low latency (67.73 msec), which can be explained by the fact that two of the 3 components are deployed on the same machine, while only the cloud's *Results Displayer* is placed in the cloud. Finally, while in test case 3, two of the components are deployed on the same node, the fact that the ML module (i.e., *Parade Footage Analyzer*) is the only component on the fog node resulted in a very high latency (~ 1s). These results could mean that the original test case chosen for this work (i.e., test case 1) is a good compromise to reduce the end-to-end latency. In particular, test case 1 is suitable even for more complicated scenarios, where computationally intensive components (compared to our simple results displayer) must be placed in the cloud.

V. CONCLUSION

This paper proposes a novel IoT PaaS architecture for NFV-based hybrid cloud/fog systems. The proposed PaaS is driven by two IoT scenarios; a smart parade scenario and a smart accident management scenario. The proposed PaaS architecture automates the provisioning of IoT applications over cloud and fog resources. In contrast to the existing IoT PaaS solutions, the proposed solution enables the discovery of existing cloud and fog nodes as well as the generation of application graphs with different sub-structures (e.g., selection, parallel). The proposed PaaS architecture is implemented as a Proof-of-Concept prototype for a smart parade scenario, and a set of experiments are conducted to evaluate the feasibility of the architecture. The results show the higher latency of executing the deployment plan compared to the migration plan. In addition, the end-to-end latency was analyzed over three different test cases with a different distribution of the application components over the cloud and the fog nodes. The performance of distributed and centralized PaaS was also analyzed considering the placement of PaaS modules in clouds and fogs in different geographical locations. The results show that the PaaS needs an efficient placement algorithm for its modules as well as for the application components in order to obtain optimal results in terms of latency.

ACKNOWLEDGMENT

This work is partially funded by the CISCO CRC program (Grant #973107), the Canada Research Chair Program, and the Canadian Natural Sciences and Engineering Research Council (NSERC) through the Discovery Grant program.

REFERENCES

[1] L. M. Vaquero, L. Rodero-Merino, J. Caceres, and M. Lindner, "A Break in the Clouds: Towards a Cloud Definition," *SIGCOMM Comput Commun Rev*, vol. 39, no. 1, pp. 50–55, Dec. 2008.

[2] P. P. Ray, "A survey on Internet of Things architectures," *J. King Saud Univ. - Comput. Inf. Sci.*, vol. 30, no. 3, pp. 291–319, Jul. 2018, doi: 10.1016/j.jksuci.2016.10.003.

[3] C. Mouradian, D. Naboulsi, S. Yangui, R. H. Glitho, M. J. Morrow, and P. A. Polakos, "A Comprehensive Survey on Fog Computing: State-of-the-Art and Research Challenges," *IEEE Commun. Surv. Tutor.*, vol. 20, no. 1, pp. 416–464, Firstquarter 2018.

[4] S. Yangui *et al.*, "A platform as-a-service for hybrid cloud/fog environments," in *2016 IEEE International Symposium on Local and Metropolitan Area Networks (LANMAN)*, 2016, pp. 1–7.

[5] M. Liyanage, C. Chang, and S. N. Srirama, "mePaaS: Mobile-Embedded Platform as a Service for Distributing Fog Computing to Edge Nodes," in *2016 17th International Conference on Parallel and Distributed Computing, Applications and Technologies (PDCAT)*, 2016, pp. 73–80.

[6] C. Mouradian, S. Kianpisheh, M. Abu-Lebdeh, F. Ebrahimnezhad, N. Tahghigh Jahromi, and R. H. Glitho, "Application Component Placement in NFV-based Hybrid Cloud/Fog Systems with Mobile Fog Nodes," *IEEE J. Sel. Areas Commun.*, vol. 37, no. 5, pp. 1130–1143, May 2019.

[7] R. Mijumbi, J. Serrat, J. L. Gorricho, N. Bouten, F. D. Turck, and R. Boutaba, "Network Function Virtualization: State-of-the-Art and Research Challenges," *IEEE Commun. Surv. Tutor.*, vol. 18, no. 1, pp. 236–262, Firstquarter 2016.

[8] ETSI, "Network Functions Virtualisation (NFV); Architectural Framework." ETSI GS NFV 002 V1.1.1, Dec-2013.

[9] P. Gora and I. Rüb, "Traffic Models for Self-driving Connected Cars," *Transp. Res. Procedia*, vol. 14, pp. 2207–2216, Jan. 2016.

[10] C. Pahl, S. Helmer, L. Miori, J. Sanin, and B. Lee, "A Container-Based Edge Cloud PaaS Architecture Based on Raspberry Pi Clusters," in *2016 IEEE 4th International Conference on Future Internet of Things and Cloud Workshops (FiCloudW)*, 2016, pp. 117–124.

[11] E. Yigitoglu, M. Mohamed, L. Liu, and H. Ludwig, "Foggy: A Framework for Continuous Automated IoT Application Deployment in Fog Computing," in *2017 IEEE International Conference on AI Mobile Services (AIMS)*, 2017, pp. 38–45.

[12] E. Saurez, K. Hong, D. Lillethun, U. Ramachandran, and B. Ottenwälder, "Incremental Deployment and Migration of Geo-distributed Situation Awareness Applications in the Fog," in *Proceedings of the 10th ACM International Conference on Distributed and Event-based Systems*, New York, NY, USA, 2016, pp. 258–269.

[13] B. Donassolo, I. Fajjari, A. Legrand, and P. Mertikopoulos, "Fog Based Framework for IoT Service Provisioning," in *2019 16th IEEE Annual Consumer Communications Networking Conference (CCNC)*, 2019, pp. 1–6.

[14] M. Tao, K. Ota, and M. Dong, "Foud: Integrating Fog and Cloud for 5G-Enabled V2G Networks," *IEEE Netw.*, vol. 31, no. 2, pp. 8–13, Mar. 2017.

[15] S. Tuli, R. Mahmud, S. Tuli, and R. Buyya, "FogBus: A Blockchain-based Lightweight Framework for Edge and Fog Computing," *J. Syst. Softw.*, vol. 154, pp. 22–36, Aug. 2019.

[16] Y. Liu, J. E. Fieldsend, and G. Min, "A Framework of Fog Computing: Architecture, Challenges, and Optimization," *IEEE Access*, vol. 5, pp. 25445–25454, 2017.

[17] S. N. Afrasiabi, S. Kianpisheh, C. Mouradian, R. H. Glitho, and A. Moghe, "Application Components Migration in NFV-based Hybrid Cloud/Fog Systems," *ArXiv190600749 Cs*, May 2019.